\documentclass[prl,aps,singlecolumn,superscriptaddress,preprintnumbers,showpacs,tightenlines,floatfix,amsmath,amssymb]{revtex4}

\usepackage{bm}
\usepackage{amsmath}
\usepackage{amssymb}
\usepackage{amsxtra}
\usepackage{amscd}
\usepackage{amsthm}
\usepackage{amsfonts}
\usepackage{eucal}
\usepackage{latexsym,amsthm}

\newcommand{\nc}{\newcommand}
\nc{\on}{\operatorname}
\nc{\wt}{\widetilde}
\nc{\Wick}{{\mathbb :}}
\nc{\R}{{\mathbb R}}

\newcommand{\beq}{\begin{equation}}
\newcommand{\eeq}{\end{equation}}
\newcommand{\bmul}{\begin{multline}}
\newcommand{\emul}{{\end{multline}}}
\newcommand\beqa{\begin{eqnarray}}
\newcommand\eeqa{\end{eqnarray}}
\newcommand\bea{\begin{array}}
\newcommand\eea{\end{array}}
\newcommand\ba{\begin{array}}
\newcommand\ea{\end{array}}

\newcommand{\neqa}{\nonumber\end{eqnarray}}

\newcommand{\eq}[1]{eq.(\ref{#1})}

\newcommand{\ur}[1]{(\ref{#1})}

\renewcommand{\d}{\partial}
\renewcommand{\O}{{\cal O}}

\newcommand{\<}{{\langle}}
\renewcommand{\>}{{\rangle}}

\newcommand{\cA}{{\cal A}}
\nc{\CH}{{\mathcal H}}

\nc{\CM}{{\mathcal M}}
\nc{\CN}{{\mathcal N}}

\newcommand{\re}{\relax{\rm I\kern-.18em R}}

\renewcommand{\)}{\right)}
\renewcommand{\(}{\left(}
\nc{\al}{{\alpha}}
\nc\comment[1]{}

\def\eps{{\epsilon}}
\def\cJ{{ J}}

\begin{document}
\title{Higgs effect in consistent Kaluza-Klein models with branes}

\author{Sergey Slizovskiy}
\email{Sergey.Slizovskiy@fysast.uu.se}
\affiliation{Department of Physics and Astronomy, Uppsala University,
P.O. Box 803, S-75108, Uppsala, Sweden}

\preprint{UUITP-11/10}

\begin{abstract}
We continue consideration of models where the Higgs effect is produced by the presence of 3-brane fluctuating in compact extra dimensions.
The consistent examples of such models may be obtained from previously known solutions of 6D supergravity. The low-energy limit for these
models coincides with a version of Higgs model written in terms of gauge-invariant supercurrent variables.

We also elaborate on vortices in the Abelian version of the theory and show how vortices that are singular from the 4D point of view
become allowed due to higher-dimensional resolution of singularity.
\end{abstract}


\date{\today}

\maketitle

\section{Introduction}
The idea of geometrization of physical phenomena has been very appealing since long ago. One of the most prominent ideas of this
type is the Kaluza-Klein idea that unifies gauge interactions with gravity \cite{KK}. Although not in it's original simple forms, this idea is still 
alive as a part of various moduli stabilization scenarios. Much recent work was devoted to finding consistent supergravity solutions
and studying their stability (see e.g. \cite{GGP, susha, susha old}). Some of these solutions look very close to those  studied naively in the old days. 
These were claimed to give a non-Abelian Yang-Mills theory in the low energy limit (assuming the dilaton stabilization).  

In the work \cite{HiggsKK} we  proposed a framework in which both the Higgs field and the non-Abelian gauge field have
geometrical origin. Namely, consider  a  Kaluza-Klein type model with 3+1 dimensional brane stretched along the  space-time.   
Then the gauge group ``direction'' of the Higgs field corresponds to the brane position  while it's modulus is related the mode corresponding to the radius of the sphere.
It is interesting that such splitting into ``angular'' and radial parts arises naturally in gauge theories in the framework inspired by spin-charge separation type techniques.
Namely, the phenomenology of the resulting low-energy theories is described in terms of gauge-invarint non-Abelian supercurrents \cite{CFN,CN2007}.  
In the previous work \cite{HiggsKK}  we studied this scenario on the simple examples of old-fashioned models, which were not a consistent truncation of gravity.

In this note we show that similar scenario can be realized in the model, which is a consistent truncation of 6D supergravity. This gives our
considerations a firm theoretical ground and also leads to conjectures the relation of Higgs mass to dilaton potential that is expected to be
computed from string theory.    

Apart from that, we continue the studies of distinguishing features of the proposed scenario. Namely, 
we show that monopoles and vortices are expected to have a different core structure in Kaluza-Klein theories with branes,  namely,
the brane my wind around extra dimensions in the center of a vortex.     
 
\section{Rugby ball solution for extra dimensions with branes}  
Consider the framework of 6D N=1 supergravity, see e.g. \cite{GGP, susha} for notations.
 Consider an explicit solution of 6D supergravity, found by  Gibbons, G\"uven and Pope  \cite{GGP}, given by
\beqa
ds^2 = e^{A(u)} (\eta_{\mu\nu} dx^\mu dx^\nu + du^2)  + e^{B(u)} \frac{r_0^2}{4} d \varphi^2 \\
e^{\phi/2} = e^A = \sqrt{\frac{f_1}{f_0}}  e^{\phi_0/2}, \  \ e^B = 4 \alpha^2 e^A \frac{\cot^2(u/r_0)}{f_1^2}  \\
\cA = -\frac{4 \alpha}{q \kappa f_1} Q \, d \varphi \\
H_{MNP} = 0
\eeqa
with $\kappa$ being a 6D gravitational constant, $q$ and $\alpha$ being generic real parameters, and $Q$ being a generator of $U(1)$ subgroup of the gauge group. 
The variable $u$:  $0\leq  u \leq \bar u  = \pi r_0/2 $ parameterizes together with $\varphi$ the compact extra dimension, 
\beqa
f_0 = 1 + \cot^2\(\frac{u}{r_0}\)  , \  \  \  \ f_1 = 1+\frac{r_0^2}{r_1^2}  \cot^2 \(\frac{u}{r_0} \)
\eeqa
The constant $\phi_0$ is a continuous classical parameter of the solution,  but the action on the solution depends on $\phi_0$ as $e^{\phi_0}$ \cite{susha}. 
The field $\cA$ is a collection of gauge fields with couplings $g$ , including, in particular, the $U(1)_R$ subgroup with coupling $g_1$.  $H_{MNP}$ is a Kalb-Ramond gauge field strength and
$\phi$ is a dilaton field. 
  
 This solution is supported by two branes located at $u=0$ and $u=\bar u$. The gauge field $\cA$ depends on the brane positions and obeys
 the Dirac quantization condition $-e \alpha \frac{r_1 \, g}{r_0 \, g_1} = N \in \mathbb Z$. $N$ is called the monopole charge 
 and the solution was found in \cite{susha} to be marginally stable for $N = \pm 1$.
Thus $r_1/r_0$ is also fixed, since  $\alpha$ is fixed by choosing brane 
 tensions (see below).   Near the branes the metric is that of a cone
 with deficit angles 
 \beq
 \delta = 2 \pi \(1-|\alpha| \frac{r_1^2}{r_0^2} \) \  \  \  \mbox{and} \ \ \ \bar \delta = 2 \pi \(1-|\alpha|  \)   
 \eeq    
 and brane tensions are $T = 2 \delta/\kappa^2$ and $\bar T = 2 \bar \delta/\kappa^2$. The conical singularities are sourced by
 brane tensions.   The parameter $r_0$  of this solution is given by $r_0^2 = \kappa^2/(2 g_1^2) e^{\phi_0/2}$,  and $r_1^2 = 8/q^2$ with $q$ being a generic real number.   
Thus the radius of the compact dimension in combination together with the dilaton forms a classically flat direction.  This is called the dilaton run-away problem. 
It is argued in several works
that this flat direction is lifted by a non-perturbative potential: It was proposed to be stabilized by accounting 
for the RG improvement of K\"ahler potential in the model without brane sources the work \cite{susha old}.
In the work \cite{Choi} a stabilization of dilaton mode was also proposed.
In the work \cite{cap reg} an alternative mechanism for stabilization based on resolving the ``thick'' brane structure was proposed.  
We will not be specific about the exact form of dilaton-stabilization potential in this paper.

Due to non-trivial warp-factor $A$ here, we restrict ourselves to considering the case $r_0 = r_1$ in which case $A=0$,  the background value of the dilaton field is zero $\phi = 0$ 
 and  the tensions of two branes
are equal: $T = \bar T$. This is called the  unwarped ``rugbyball'' compactification. It is consistent with Kaluza-Klein ansatz and the metric reduces to 
\beqa
ds^2 = \eta_{\mu\nu} dx^\mu dx^\nu  +  \frac{r_0^2}{4}\(d \theta^2 + \alpha^2 \sin^2\theta  d \varphi^2 \)
\eeqa  
where $\theta = 2 u /r_0$.
Note that this is also a solution of non-supersymmetric 6D Einstein--Yang-Mills model with cosmological term (EYM$\rm\Lambda$). Similar solutions were studied earlier in
\cite{Carroll,Navarro}.

Consider the case when $\delta \ll 1$. It corresponds to brane tensions being small in Kaluza-Klein scale and the solution being almost a sphere of radius $r=r_0/2$. The two branes of equal tension sit in the two poles of the sphere.
The full spectrum for fluctuations around ``rugbyball'' solution was considered in \cite{susha}, where also a stability analysis was performed. 
In the analysis of the paper \cite{susha} the choice of the light-cone gauge simplified the calculations since tensor, vector and scalar fluctuations decoupled in this gauge. 
The direct analysis of brane fluctuations was not completed there due to need to consider thick brane to get rid of singularities.   
But that is not needed for our purpose of getting the low-energy modes, since the  Kaluza-Klein gauge transformation (reparametrization) transfers the low-energy brane 
fluctuations into the low-energy vector fluctuations.

We are interested in modes that are massless or with masses
proportional to the brane tension, their spectrum, extracted from \cite{susha}, is actually the spectrum
of physical fluctuations of gauge-invariant supercurrents, that consist of branons (brane fluctuation modes) together with vector fields.

The relevant part of the spectrum for vector components of gravitons mixed with other vector fields  is given by 
$M_{nm}^2 = \frac{4}{r_0^2} (1 + \frac{r_0^2}{4} \mu^2_{n m } - \sqrt{1+r_0^2 \mu^2_{nm}})$ with
$\mu^2_{nm} = \frac{4}{r_0^2} (n+ |m| \omega) (n+ |m| \omega +1 ) $ with $n\in{ \mathbb Z}_+$,  $m\in {\mathbb Z} $ and $\omega = \frac1{1- \delta/2 \pi}$.
From here we extract the mode with $\{n,m\} = \{1,0\}$ which is exactly massless and corresponds to the unbroken $U(1)$ axial symmetry and 
two modes with $\{n,m\} = \{0,\pm 1\}$ with masses 
\beq
M^2_{\{0,\pm 1\}} = \frac{2 \delta}{\pi r_0^2} + \O(\delta^2) =  \frac{T \kappa^2}{\pi r_0^2} + \O(\delta^2)  \approx  16 \pi G T 
\eeq  
where we converted to units of 4D gravitational constant using $1/(16 \pi G) = 4 \pi r^2 / \kappa^2 $, which is correct in the asymptotic vacuum region, where the radius $r$ 
is expected to be a constant, minimizing the dilaton-stabilization potential.
For comparison, the same mass in the 6D Einstein-Yang-Mills system with cosmological constant is 
\beq
M^2_{\{0,\pm 1\} (\mathrm{ EYM \Lambda}) } = \frac{3 \delta}{\pi r_0^2} + \O(\delta^2) =  \frac{3 T \kappa^2}{2 \pi r_0^2} + \O(\delta^2)  \approx  24 \pi G T
\eeq  
 Note that in the supergravity
the Kaluza-Klein gauge field in made out of a linear combination of vector component of gravity field with gauge field $\cA$. 

This is to be compared with the naive result for the Higgs field mass, that we can obtained accounting only for changing of the brane world-volume due to brane fluctuations  
and not accounting for changing of gravity
and gauge fields, caused by moving source, see \cite{HiggsKK}. 

Let us briefly review it here, specializing to the case of compact manifold being $S^2 = SU(2)/U(1)$.
We have three Killing vectors on $S^2$:
 \beqa \label{KillingS2}
 K^1 = -\sin \varphi \d_\theta - \cot \theta \sin \varphi \d_\varphi ; \ \ K^2 = \cos \varphi \d_\theta  - \cot \theta \sin \varphi \d_\varphi   
\ ;  \ K^3 = \d_\varphi  
\eeqa
and define two massive gauge-invariant supercurrents:
\beq 
 \cJ^m_\mu =  \frac{\d X^m}{\d x^\mu} + K^{a m}(X) A^a_\mu
\eeq
where $X(x)$ denotes the compact coordinates $X^m$ of the 3-brane as a function of 4D coordinates $x_\mu$; the indices $a=1,2,3$ label the gauge group generators and
$m,n = (\theta,\varphi)$ label the compact space coordinates.  
  Then the effective Kaluza-Klein action reads
\beq \label{naive result}
S_{YM} =   \int d^4 x \sqrt{-g_{4D}} \left \{ - \frac{k V_{int} r ^2}{\kappa^2}  \frac14 (F_{\mu\nu})^2 +    \frac{1}{2} T g^{\mu \nu} \cJ^m_\mu g_{mn}  \cJ^n_\nu  \right \}
\eeq 
 with $V_{int} = 4 \pi r^2 $ being a volume of compact dimension and  $k =  \dim(G/H)/ dim G=  \dim(SU(2)/U(1))/ \dim SU(2)   = 2/3$ and $\kappa$ being a 
 6D gravitational coupling  \cite{Salam, HiggsKK}.

It is important to observe that we get at least qualitatively correct result here by discarding the ''vacuum'' part of the 3-brane action: $\int d^4 x \sqrt {- g_{4D}} T$  which
naively could have led to a very large cosmological constant and a strongly curved 4D space, instead of the Minkowski one. Most of the brane tension goes
exclusively to curving the extra dimensions!    

Let us show how to explicitly match this result to the standard action of $SU(2)$ adjoint Higgs  Georgi-Glashow model.
Introducing a change of variables for the adjoint Higgs field: $\Phi^a = \rho \, n^a$ where $n^a$ is a unit 3D vector we may rewrite the Higgs part of the action as:
\beq
(D \Phi)^2 =  (\d_\mu \Phi^a + \epsilon^{abc}  A_\mu^b \Phi^c)^2 = \rho^2 (\d_\mu n^a + \epsilon^{abc}  A_\mu^b n^c)^2 + (\d_\mu \rho)^2
\eeq  
The kinetic and potential terms for $\rho$ are related in general to the dilaton stabilization problem and are beyond our current analysis,
we will simply make some assumptions about them. The term with $\vec n$ can be exactly matched to supercurrents.
  Indeed, choosing $\vec n = (\sin \theta \cos \varphi , \sin \theta \sin \varphi, \cos \theta)$ to be a vector pointing to the brane position ($(\theta, \varphi) = X(x)$) 
  and using the relations   
\beqa
g_{mn} = r^2 \frac{\d n^a}{\d x^m} \frac{\d n^a}{\d x^n} \\ 
\frac{\d n^b}{\d x^m} K^{c m}  =  \eps^{abc } n^a 
\eeqa
we identify
\beq
(\d_\mu n^i + \epsilon^{ijk}  A_\mu^j n^k) = \frac{\d n^i}{\d x^m} J^m_\mu   
\eeq
and 
\beq
 g^{\mu \nu} \cJ^m_\mu g_{mn}  \cJ^n_\nu =  (\d_\mu n^a + \epsilon^{abc}  A_\mu^b n^c)^2
\eeq
Also note that in the unitary gauge $\vec n = (0,0,1)$ and $ \rho^2 (\d_\mu n^i + \epsilon^{ijk}  A_\mu^j n^k)^2 = \rho^2 \((A^1)^2 + (A^2)^2\) $.

We see from \eq{naive result} that the  coefficient in front of $(F_{\mu\nu})^2$ depends on $r$ stronger than the mass term, so, changing the normalization 
of the gauge fields, to get a conventional YM kinetic term, we get that $\rho$ is inversely proportional
to $r$.  With relation
$r=r_0/2$  and doubled brane tension due to two branes we get
\beq
\<\rho^2 \> = M^2_{\{0,\pm 1\} (\text{branes only})}  =  \frac{2 T \kappa^2}{(2/3) 4 \pi r^2}  = \frac{3 T \kappa^2}{\pi r_0^2}  \approx  48 \pi G T  
\eeq

As mentioned above, the actual value of $\rho$ and the steepness of a potential for it is determined by a non-perturbative dilaton potential in the supergravity model.

It is interesting to observe that the value of the Higgs field modulus $\rho$ depends on the radius of compact dimension as 
\beq \label{inverse relation}
\rho^2 \sim \frac{T \kappa^2}{r^2}
\eeq
This relation looks somewhat counterintuitive and tells that compact extra dimensions blow-up at points where the modulus of Higgs field
vanishes. The points where the Higgs fields vanishes are commonly expected in the center of topological solutions such as vortices and monopoles. 
But we note an interesting feature that distinguishes the brane model from the conventional Higgs model:  it is allowed to have topological solutions {\it without
vanishing of the Higgs field in the center}. We show it in the next section.  

\section{Vortices and Monopoles}
We show here that the Kaluza-Klein theory with brane sources admits topological defects without vanishing of Higgs field in the center. 
For simplicity, we assume that the Higgs potential is so steep, that the modulus of the Higgs field is fixed.  This limit is opposite to  Bogomol'niy limit.
 
 We start from the $U(1)$ case, giving effectively the Abelian Higgs model. Let us search for a static solution -- it is a vortex in 3D space, where the phase of complex Higgs 
field winds on the $S^1$ at infinity.   So, consider a 2D plane orthogonal to the vortex and let us find a solution. It would mainly
consist of vacuum Higgs field, since the $U(1)$ gauge field is massive. In the standard action of Abelian Higgs model the 
configuration 
\beq \label{vortex}
\Phi(r,\phi) = \rho \, e^{i \phi}
\eeq
 has logarithmically divergent action at the core and to cure that one needs to have $\rho(0)=0$.
That costs much energy if the Higgs potential is steep. 

The basic observation of this section is that in the effective Higgs model that arises from brane such configuration does not have a divergent action.
Consider the radius of one extra dimension to be fixed to $r_0$.  The gauge-invarint supercurrent reads: 
$
  J_\mu = A_\mu + \d_\mu x^5(x) 
$
  where $x^5(x)$ is a coordinate of a 3-brane, it is identified with $\phi$.  
The total action reads
\beq
S = \frac{1}{16 \pi G} \int d^4 x \sqrt{-g} (R + r_0^2 (F_{\mu\nu})^2)  + T \int d^4 x \sqrt g  \sqrt {\(1 +   r_0^2 J_\mu J^\mu  \)} 
\eeq
Recall that to obtain a standard Higgs model we made a derivative expansion of the brane volume, assuming that the brane
fluctuates slowly on the scales or compact dimension ($J \ll r$) \cite{HiggsKK}:
$    \sqrt {\(1 +   r_0^2 J_\mu J^\mu  \)} \approx 1 + \frac12 r_0^2  J^2 + ...  $.  
But the assumption $J \ll r$ breaks down near the center of the vortex, so, one needs to consider the full action $\sim \sqrt {\(1 +   r_0^2 J_\mu J^\mu  \)} - 1$
if the brane model is to be taken seriously. 
If we consider geometrically the model where the phase of the Higgs field corresponds to a brane position in the compact $S^1$
then near the vortex line the brane looks as a spiral staircase and at the vortex line the brane fills the compact $S^1$.  
It is not hard to compute the area of the brane for the geometry corresponding to a vortex \ur{vortex}: take $A_\mu = 0$ and $J_\mu = \d_\mu \phi$ where
$\phi$ is a polar angle, then
\beq
 M_{vortex}  = T \int_{x_0 = const}  (\sqrt g_{induced} - \sqrt g)  = T l \int_0^{2 \pi}  d \phi \int_0^R d r (\sqrt{r^2+ r_0^2} - r ) = T l \pi r_0^2 \, {\rm arcsh}(R/r_0) 
\eeq  
  here $(r,\phi)$ are polar coordinates on the plane orthogonal to vortex line and $r_0$ is a  Kaluza-Klein radius of compact $S^1$ and
  $R$ is a cut-off at infinity to regulate the IR divergence (in practice it should be regulated by considering a dipole or a ring).
  The length of the vortex is $l$, but this parameter is formal since if we assume that the brane cannot have holes in it, then the winding number is conserved
  and vortex cannot have boundaries.  That means that vortices should either exist forever or be created and annihilated in pairs, as usual. 
   To convert the mass to standard units we note that $r_0^2 = \frac{16 \pi G}{e^2}$
 and  the mass of the gauge field is $M_A^2 =  16 \pi G T$. 
 So, the mass per unit length of the vortex is
 \beq
 M_{vortex}/l = \pi \frac{M_A^2}{e^2}  {\rm arcsh}(R/r_0)   
 \eeq

There can be also singular lines of gauge field $A_\mu$ in the Abelian Higgs model, but these occur to be the same configurations as singular
lines of $\Phi$. Indeed, if we write in 2D plane orthogonal to vortex line $A = A_r dr + A_\phi d \phi$ then at $r=0$ the regular field must have 
$\left. A_\phi \right |_{r=0} = 0$.  If we have instead $A_\phi = 1$ then it is gauge-equivalent to winding of $\Phi$ at $r=0$, so, it is the same
vortex. 
   
   Note that if we expand the action in $r$, as
we did to get a polynomial Higgs action, all terms would give divergent integrals.  
   
   For non-Abelian case we can also consider the standard monopole solutions and evaluate the consequences of keeping the exact action for
the supercurrents.   This effect is small and depends on the detailed form of the Higgs potential.  

 {\it Acknowledgements} I am grateful to Antti Niemi for drawing my interest to the subject  and to Susha Parameswaran and Thomas Van Riet for clarifying discussions. The work was supported by   STINT Institutional Grant and VR Grant 2006-3376.

\end{document}